\documentclass[amsmath,amssymb,onecolumn,superscriptaddress]{revtex4}

\usepackage{graphicx}% 
\usepackage{dcolumn}% 
\def\be{\begin{equation}}
\def\ee{\end{equation}}
\def\beq{\begin{eqnarray}}
\def\eeq{\end{eqnarray}}

\usepackage{bm}
\usepackage{graphicx}
\usepackage{dcolumn}
\usepackage{amsmath}
\usepackage[latin1]{inputenc}
\usepackage{graphicx, psfrag}
\usepackage{amssymb}
\usepackage[colorlinks=true, citecolor=blue, urlcolor = blue, linkcolor= red, bookmarks=true]{hyperref}
\usepackage{float}
\usepackage{amsmath}
\usepackage{amsfonts}
\usepackage{dcolumn}
\usepackage{hyperref}
\usepackage{subfigure}
\usepackage{pgfplots}
\usepackage{epstopdf}
\usepackage{booktabs}
\usepackage{orcidlink}
\usepackage{xcolor}

\begin{document}
\title{Relativistic isotropic stellar model in $f(R,\,T)$ gravity with Durgapal- IV Metric}

\author{Pramit Rej \orcidlink{0000-0001-5359-0655}\footnote{Corresponding author}}
\email{pramitrej@gmail.com, pramitr@sccollegednk.ac.in }
 \affiliation{Department of Mathematics, Sarat Centenary College, Dhaniakhali, Hooghly, West Bengal 712 302, India}

\author{Piyali Bhar \orcidlink{0000-0001-9747-1009}}
\email{piyalibhar90@gmail.com, piyalibhar@associates.iucaa.in}
\affiliation{Department of Mathematics, Government General Degree College Singur, Hooghly, West Bengal 712 409, India}

\begin{abstract} \noindent
In this work, a new static, non-singular, spherically symmetric fluid model has been obtained in the background of $f(R,\,T)$ gravity. Here we consider the isotropic metric potentials of Durgapal-IV [M.C. Durgapal, J. Phys. A  {\bf 15} 2637 (1982)] solution as input to handle the Einstein field equations in $f(R,\,T)$ environment. For different coupling parameter values of $\chi$, graphical representations of the physical parameters have been demonstrated to describe the analytical results more clearly. It should be highlighted that the results of General Relativity (GR) are given by $\chi=0$. With the use of both analytical discussion and graphical illustrations, a thorough comparison of our results with the GR outcomes is also covered. The numerical values of the various physical attributes have been given for various coupling parameter $\chi$ values in order to discuss the impact of this parameter. Here we apply our solution by considering the compact star candidate LMC X-4 [M.L. Rawls et al., Astrophys. J. {\bf 730} 25 (2011)] with mass$=(1.04 \pm 0.09)M_{\odot}$ and radius $= 8.301_{-0.2}^{+0.2}$ km. respectively, to analyze both analytically and graphically. To confirm the physical acceptance of our model, we discuss certain physical properties of our obtained solution such as energy conditions, causality, hydrostatic equilibrium through a modified Tolman-Oppenheimer-Volkoff (TOV) conservation equation, pressure-density ratio, etc. Also, our solution is well-behaved and free from any singularity at the center. From our present study, it is observed that all of our obtained results fall within the physically admissible regime, indicating the viability of our model.

\end{abstract}

%\keywords{General relativity, Compact star, $f(R,T)$ gravity, Causality condition}

%\ccode{PACS numbers: 04.20.-q; 04.40.Nr; 04.40.Dg}

\maketitle

\section{Introduction}
The universe has been expanding ever since the Big Bang. After Big Bang, this expansion is fairly normal, but after some time, it is anticipated that the rate of expansion would slow. However, in 1998, two separate investigations, the High-Z Supernova Search Team, and the Supernova Cosmology Project, discovered that the universe's pace of expansion is not decreasing, but rather in an accelerated manner.
The reason for this accelerated expansion has left scientists puzzled. This suggests that the cosmos is expanding at an accelerated rate due to some kind of hidden energy. The accelerated expansion of the cosmos has been accounted for in a variety of ways up to this point, including the existence of dark energy and modified formulations of the general theory of relativity.
Among the many solutions put forth, different modified gravity theories like unimodular gravity \cite{nojiri2016unimodular, garcia2019cosmic}, $f(R)$ gravity \cite{nojiri2007introduction,nojiri2011unified}, $f(\mathcal{Q}, T )$ gravity \cite{xu2019f, arora2020f}, teleparallel gravity \cite{aldrovandi2013teleparallel, bahamonde2022teleparallel}, $f(\mathcal{G}, T )$ gravity \cite{sharif2016energy, bhatti2018role}, $f(R, \mathcal{G})$ gravity \cite{atazadeh2014energy, de2011stability} and so on have been proposed by different researchers. According to early claims, an exotic type of energy with negative pressure and positive energy density is responsible for the late-time acceleration of the expansion of the universe. However, with the right geometrical adjustments, this problem can be resolved without the use of such exotic energy.

Harko et al. \cite{Harko:2011kv} assumed a weak coupling between matter and geometry and developed a unique type of gravity they named as $f(R,\,T)$ gravity. They tried to resolve the problem of accelerated expansion of the universe using this theory. This $f(R,\,T)$ gravity provides an alternative explanation for the current cosmic acceleration without involving the construction of extra spatial dimensions or the insertion of an exotic dark energy component. Matter contents may contribute to cosmic acceleration in $f(R,\,T)$ gravity in addition to geometrical contributions to overall cosmic energy density \cite{Zubair:2015gsb}. In recent decades, the $f(R,\,T)$ theory has gained popularity among researchers. The gravitational Lagrangian in this modified theory of gravity is made up of an arbitrary function of the Ricci scalar $R$ and the trace of the stress-energy tensor $T$. As a result of the covariant divergence of the stress-energy tensor, Harko et al. developed gravitational field equations in metric formalism, as well as equations of motion for test particles. The trace component could be connected to the possibility of imperfect fluids.

Within this theory, Tangphati et al. \cite{Tangphati:2022mur} examined compact static configurations with interacting quark EoS, which is a homogeneous, neutral 3-flavor interacting quark matter with O(ms4) corrections. Bhattacharjee \cite{Bhattacharjee:2022lcs} creates feasible inflationary cosmic solutions which are consistent with the latest Planck and BICEP2/Keck Array data using mimetic $f(R,\,T)$ gravity, a Lagrange multiplier, and a mimetic potential. Baffou et al. \cite{Baffou:2021ycm} explored the cosmological inflation scenario in the framework of the $f(R,\,T)$ theory of gravity, and Prasad et al. \cite{Prasad:2021yrf} investigate the possibility of a compact star configuration model. Yousaf \cite{Yousaf:2021tol} studied axial and reflection-symmetric self-gravitating sources using a well-known Karmarkar condition to produce a well-behaved embedding class-one solution with a required linear function for $f(R,\,T)$. Using the Buchdahl {\em ansatz}, Kumar et al. \cite{Kumar:2021vqa} proposed an isotropic compact star model in $f(R,\,T)$ gravity. Sharif and Waseem \cite{Sharif:2019oni} explored the effects of charge on stellar objects, known as gravastars, under the influence of $f(R,\,T)$ gravity by taking into account the Mazur and Mottola conjecture in general relativity. A brand-new, efficient technique for building self-gravitating systems that rely on imperfect fluid distributions was presented by Maurya et al \cite{Maurya:2021aio}. By combining two geometrical schemes: gravitational decoupling via minimal geometric deformation and the embedding approach, notably, the class I grip-this method was developed within the framework of the $f(R,\,T)$ gravity theory. The $f(R,\, T)$ model occasionally exhibits more pleasant behavior than its Einstein equivalent, according to Hansraj and Banerjee \cite{Hansraj:2018jzb}, who studied the behavior of well-known stellar models in the context of the $f(R,\, T)$ modified gravity theory. Further research into this $f(R,\, T)$ theory by Barrientos et al. \cite{Barrientos:2018cnx} revealed that the resulting field equations are quite similar to their metric-affine $f(R)$ siblings after an effective energy-momentum tensor is applied to the system. Gamonal \cite{Gamonal:2020itt} chose a minimal couplin between matter and gravity in order to study the slow-roll approximation to cosmic inflation in the context of $f(R,\, T)$ gravity. A universal $5\mathcal{D}$ metric is used in Moraes \cite{Moraes:2015kka} proposes $f(R,\, T)$ theory to apply precise cosmological solutions derived from Wesson's induced matter model. Bhar and Rej \cite{bhar2022isotropic} recently studied an isotropic Buchdahl relativistic fluid sphere in the presence of $f(R,\, T)$ gravity. \par
In relativistic astrophysics, exact solutions to the Einstein field equations for the study of compact stellar structures are very crucial. Since the discovery of Einstein's gravitational field equations, relativists and astrophysicists have been attempting to create perfect fluid models of superdense objects. It is still very complicated to construct realistic compact configurations satisfying the Einstein field equations using a wide range of physical conditions. Schwarzschild \cite{schwarzschild1916gravitationsfeld} formulated the initial solution to Einstein's field equations. That proposed solution demonstrates the neighborhood of a spherically symmetric, static compact stellar object having vanishing density and pressure. This magnificent theory has been significantly elaborated in later decades. The simplest scenario in this context is the analysis of spherically symmetric, static configurations made up of ideal fluid distributions and isotropic pressure distributions (i.e., $p_r=p_t$). Several studies have been done on this topic. Professor Tolman's pioneer research works \cite{tolman1939static} provide further developments for this topic. He established a series of analytical solutions that explain stellar interiors with spherically symmetric, static geometry, similar to the matter distribution of a perfect fluid  while maintaining isotropic pressure. 
Naturally, neither neutron stars nor any other stars are totally composed of perfect fluid. However, these solutions could be used to generate a superdense object model suitable for the numerical analysis of stellar structure. For a variety of reasons, physicists and astronomers are interested in the study of relativistic fluid spheres. In this study, by taking into account the metric components of Durgapal-IV solution \cite{durgapal1982class}[or Durg IV metric in accordance with the Delgaty and Lake 1998 classification], we are curious to learn more about the physical properties of the compact stellar object LMC X-4, which depicts a static spherically symmetric configuration along with an isotropic fluid matter by solving the Einstein field equations. There are several earlier research works on Durgapal IV solution in literature \cite{Fuloria:2012yol, Mehta:2013ir, Tello-Ortiz:2020nuc, Contreras:2022vmk}. Murad and Fatema \cite{murad2013family} employed Durgapal's solution to present a new family of interior solutions of Einstein-Maxwell field equations in GR for a static spherically symmetric charged perfect fluid. Islam et al. \cite{islam2019analytical} took the Durgapal-IV metric to model strange stars analytically. Maurya and Gupta \cite{maurya2011extremization} obtained a class of charged superdense star models by considering the Durgapal type space-time metric.

In addition to discussing the stability and physical characteristics of the compact star LMC X-4, the aim of this research is to investigate the appearance of $f(R,\,T)$ gravity while simulating realistic configurations of compact stellar objects based on the Durgapal-IV metric. By utilizing a particular type of $f(R,\,T)$ gravity model, we examine several structural aspects in the setting of an isotropic matter source. We shall explore in detail the Tolman-Oppenheimer-Volkoff (TOV) equation, the mass-radius relation, the compactness parameter, the surface redshift, the stability, and different energy conditions.

The structure of this paper is represented as follows: an overview of $f(R,\,T)$ gravity in the context of isotropic matter distributions is presented in Section \ref{sec2}. Section \ref{sec3} deals with solutions of field equations using Durgapal's fourth model. Section \ref{sec4} deals with junction conditions, here we match the interior line element to Schwarzschild's exterior line element for calculating the values of the unknown constants. Section \ref{sec5} examines some physical characteristics as well as the viability of several well-known compact objects by using graphical analysis. The final section is covered with concluding remarks. Here we have adopted the geometricized unit system throughout our work (except Table~\ref{tb1}), where $G = c = 1$.

\section{Elementary field Equations}\label{sec2}
The form of Einstein-Hilbert(EH) action for $f(R,T )$ gravity,initially described by Harko {\em et al.} \cite{Harko:2011kv}, is given as,
\begin{eqnarray}\label{action}
S&=&\frac{1}{16 \pi}\int  f(R,T)\sqrt{-g} d^4 x + \int \mathcal{L}_m\sqrt{-g} d^4 x,
\end{eqnarray}
where $g = det(g_{\mu \nu}$) and $f ( R,T )$ being a general function of trace $T$ of the energy momentum tensor ($T_{\mu \nu}$)  along with Ricci scalar $R$. The $\mathcal{L}_m$ denotes the lagrangian matter density related
to the energy-momentum tensor $T_{\mu \nu}$. As per Landau and Lifshitz \cite{landau2013classical}, the stress-energy tensor of matter is defined as,
\begin{eqnarray}\label{tmu1}
T_{\mu \nu}&=&-\frac{2}{\sqrt{-g}}\frac{\partial \big(\sqrt{-g}\mathcal{L}_m\big)}{\partial g^{\mu \nu}},
\end{eqnarray}
and its trace $T=g^{\mu \nu}T_{\mu \nu}$. If the Lagrangian matter density $\mathcal{L}_m$ depends only on $g_{\mu \nu}$, not on its derivatives, Equn.(\ref{tmu1}) becomes,
\begin{eqnarray}
T_{\mu \nu}&=& g_{\mu \nu}\mathcal{L}_m-2\frac{\partial \mathcal{L}_m}{\partial g^{\mu \nu}}.
\end{eqnarray}
The Einstein field equations in modified $f(R,T)$ gravity corresponding to EH action (\ref{action}) is provided as,
\begin{eqnarray}\label{frt}
f_R(R,T)R_{\mu \nu}-\frac{1}{2}f(R,T)g_{\mu \nu}+(g_{\mu \nu }\Box-\nabla_{\mu}\nabla_{\nu})f_R(R,T)&=&8\pi T_{\mu \nu}-f_T(R,T)T_{\mu \nu}\nonumber\\&&-f_T(R,T)\Theta_{\mu \nu}.
\end{eqnarray}
where, $f_R(R,T)=\frac{\partial f(R,T)}{\partial R},~f_T(R,T)=\frac{\partial f(R,T)}{\partial T}$. $\nabla_{\nu}$ represents the covariant derivative
associated with the Levi-Civita connection of $g_{\mu \nu}$, $\Theta_{\mu \nu}=g^{\alpha \beta}\frac{\delta T_{\alpha \beta}}{\delta g^{\mu \nu}}$ and
$\Box \equiv \frac{1}{\sqrt{-g}}\partial_{\mu}(\sqrt{-g}g^{\mu \nu}\partial_{\nu})$ represents the D'Alembert operator.\\

Now the divergence of $T_{\mu \nu}$ can be obtained by the taking covariant divergence of (\ref{frt}) (For details see references \cite{Harko:2011kv} and\cite{koivisto2006note}) we obtain,
\begin{eqnarray}\label{conservation}
\nabla^{\mu}T_{\mu \nu}&=&\frac{f_T(R,T)}{8\pi-f_T(R,T)}\left[(T_{\mu \nu}+\Theta_{\mu \nu})\nabla^{\mu}\ln f_T(R,T)+\nabla^{\mu}\Theta_{\mu \nu}\right].
\end{eqnarray}

From equation (\ref{conservation}), we can easily verify that $\nabla^{\mu}T_{\mu \nu}\neq 0$ if $f_T(R,T)\neq 0.$ So the system will not be conserved like Einstein gravity. It is to be noted that when $f(R, T)=f(R)$, the Eqn. (\ref{frt}) reduces to the Einstein field equations of $f(R)$ gravity.\par

In curvature coordinates $(t,r,\theta,\phi)$,
\begin{equation}\label{line}
ds^{2}=-e^{\nu (r)}dt^{2}+e^{\lambda (r)}dr^{2}+r^{2}d\Omega^{2},
\end{equation}
provides the interior spacetime of the static and spherically matter distribution of a compact object, where $d\Omega^{2}\equiv (\sin^{2}\theta d\phi^{2}+d\theta^{^2})$. The metric coefficients $\nu$ and $\lambda$ are functions of $r$ only.
In this work, we assume that the fluid around a compact star is perfect. Consequently, we consider the stress-energy tensor of matter for the interior of the compact object as,
\begin{eqnarray}
T_{\mu \nu}&=&(p+\rho)u_{\mu}u_{\nu}-p g_{\mu \nu},
\end{eqnarray}
where $\rho$ and $p$ stand for matter density and isotropic pressure in modified gravity respectively. $u^{\mu}$ is the fluid four velocity satisfies the equations $u^{\mu}u_{\mu}=1$ and $u^{\mu}\nabla_{\nu}u_{\mu}=0$ and the matter Lagrangian as proposed by Harko et al.\cite{Harko:2011kv} can be taken as $\mathcal{L}_m=-p$ and the expression of $\Theta_{\mu \nu}=-2T_{\mu \nu}-pg_{\mu\nu}.$\\

Let's consider a separable functional form given by,
\begin{eqnarray}
f (R, T ) = f_1(R)+f_2(T ),
\end{eqnarray}
in the context of relativistic structures to discuss the coupling effects of matter and curvature components in $f(R,\, T)$ gravity, where $f_1(R)$ and $f_2(T)$ are arbitrary functions of $R$ and $T$ respectively. By choosing several $f_1(R)$ forms and combining them linearly with $f_2(T)$ in $f(R,\,T)$ gravity, several feasible models can be built. We take into account $f_1(R)=R$ and $f_2(T)=2\chi T $ in our current model i.e., we choose
\begin{eqnarray}\label{e}
f(R,T)&=& R+2 \chi T,
\end{eqnarray}
where $\chi$ is the coupling parameter that can be determined depending upon several physical attributes of our present model.
If $\chi=0$ then the field equation (\ref{frt}) will reduce into Einstein's GR.
Using (\ref{e}) into (\ref{frt}),
the field equations in $f(R,T)$ gravity become,
\begin{eqnarray}
G_{\mu \nu}&=&8\pi T_{\mu \nu}^{\text{eff}},
\end{eqnarray}
where $G_{\mu \nu}$ is the Einstein tensor and
\begin{eqnarray}
T_{\mu \nu}^{\text{eff}}&=& T_{\mu \nu}+\frac{\chi}{8\pi}T g_{\mu \nu}+\frac{\chi}{4\pi}(T_{\mu \nu}+p g_{\mu \nu}).
\end{eqnarray}
The field equations in modified gravity can be written as,
\begin{eqnarray}
8\pi\rho^{\text{eff}}&=&\frac{\lambda'}{r}e^{-\lambda}+\frac{1}{r^{2}}(1-e^{-\lambda}),\label{f1}\\
8 \pi p^{\text{eff}}&=& \frac{1}{r^{2}}(e^{-\lambda}-1)+\frac{\nu'}{r}e^{-\lambda},\label{f2} \\
8 \pi p^{\text{eff}}&=&\frac{1}{4}e^{-\lambda}\left[2\nu''+\nu'^2-\lambda'\nu'+\frac{2}{r}(\nu'-\lambda')\right]. \label{f3}
\end{eqnarray}
 for the line element (\ref{line}),
where $\rho^{\text{eff}}$ and $p^{\text{eff}}$ are respectively the effective energy density and effective pressure in Einstein Gravity related to the energy density ($\rho$) and pressure ($p$) and
\begin{eqnarray}
\rho^{\text{eff}}&=& \rho+\frac{\chi}{8\pi}(3 \rho-p),\label{r1}\\
p^{\text{eff}}&=& p-\frac{\chi}{8\pi}(\rho-3p),\label{r2}
\end{eqnarray}
the prime sign ($'$) indicates differentiation with respect to `r'. Using Eqs. (\ref{f1})-(\ref{f3}), we get,
\begin{eqnarray}\label{con}
\frac{\nu'}{2}(\rho+p)+\frac{dp}{dr}&=&\frac{\chi}{8\pi+2\chi}(p'-\rho').
\end{eqnarray}
In Eqn.(\ref{con}), for $\chi =0$ we obtain the conservation equation in Einstein gravity. In the next section, we shall solve the Eqns. (\ref{f1})-(\ref{f3}) to obtain the model of compact star in $f(R,\,T)$ gravity.

\section{Exact Solution employing Durgapal-IV Metric}\label{sec3}
%In eqns. (\ref{f1})-(\ref{f3}), we have three eqns. with four unknown. So we have to choose any one of them to make the system solvable. Now by our knowledge of algebra, we can choose it in $4C_1=4$ ways.\\
In this section of this paper, we want to obtain a feasible model of a compact object by solving the system of Eqns.(\ref{f1})-(\ref{f3}). For this motive, we utilize the well-known Durgapal-IV space-time metric {\em ansatz} \cite{durgapal1982class} to the static Einstein equations with a perfect fluid source given by,
\begin{eqnarray}\label{elambda}
e^{\lambda}&=&  \Bigg[\frac{C K r^2}{(1 + C r^2)^2 (1 + 5 C r^2)^{2/5}} + \frac{7 - 10 C r^2 - C^2 r^4}{7 (1 + C r^2)^2}\Bigg]^{-1}              ,
\end{eqnarray}
and,
\begin{eqnarray}\label{enu}
e^{\nu}&=& A (1 + C r^2)^4,
\end{eqnarray}
where $A$ and $K$ are both dimensionless constants and $C$ is a constant having dimension of $length^{-2}$. $A$ and $K$ can be obtained from the matching conditions as functions of $C$. \\
In the study of compact astrophysical objects, the presence of physical and geometric singularities within the star is considered one of the most vital facets. We examine the behavior of both metric potentials to check their singularity if exists. Inside the compact object, the metric potentials should be free of singularity, positive, monotonically increasing, and regular for the physical viability and stability of the model.\\
At the stellar core, $e^{\lambda}=1$ and $e^{\nu}=A$,
  and their derivatives are given by,
  \begin{eqnarray}
  (e^{\lambda})'&=& -\frac{14 C r (1 + C r^2) \bigg\{8 (-3 + C r^2) (1 + 5 C r^2)^{7/5} +
   7 K \bigg(1 + C r^2 (2 - 7 C r^2)\bigg)\bigg\}}{(1 + 5 C r^2)^{3/5} \bigg\{-7 (1 + 5 C r^2)^{2/5} +
   C r^2 \bigg(-7 K + (10 + C r^2) (1 + 5 C r^2)^{2/5}\bigg)\bigg\}^2},\\
  (e^{\nu})'&=& 8 A C r (1 + C r^2)^3,
 \end{eqnarray}
\\
  The derivatives of the metric coefficients vanish at the center of the star, indicating that the metric coefficients are regular at the center of the star. \begin{figure}[htbp]
    \centering
        \includegraphics[scale=.55]{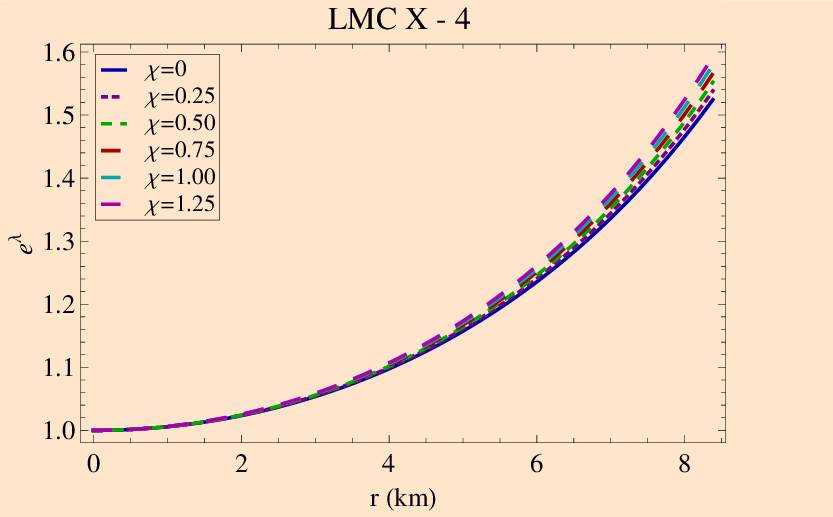}
        \includegraphics[scale=.55]{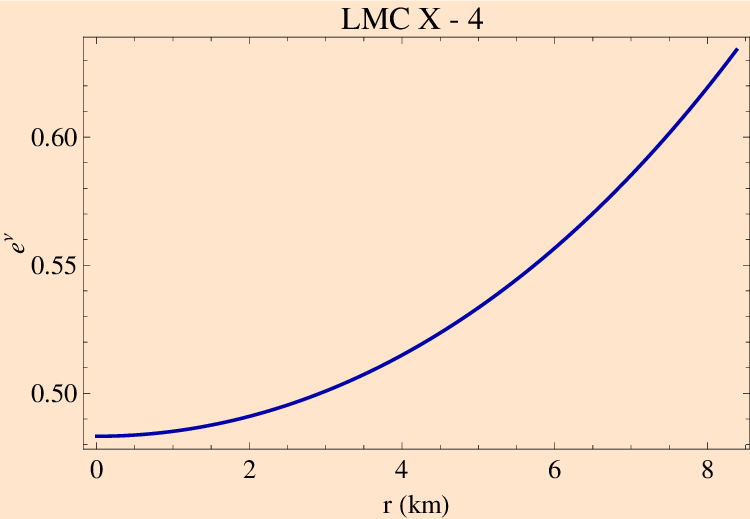}
       \caption{$e^{\lambda}$ and $e^{\nu}$ are shown against `r'.}\label{metric}
\end{figure}
Fig.~\ref{metric} depicts the characteristics of metric coefficients. Both the metric potentials are found to be consistent with the previously mentioned conditions. The graphical behavior shows that the value of both metric potentials is minimum at the center, then increases nonlinearly until it attains its maximum at the boundary surface.\par
To explore the entire structure of stellar models within $f(R,\, T)$ gravity, we must obtain first the expressions for physical parameters such as effective matter density and isotropic pressure. The effective matter density and pressure are
evaluated by using the expressions of metric potentials as,
\begin{eqnarray}
\rho^{\text{eff}}&=&\frac{C \bigg \{8 (1 + 5 C r^2)^{7/5} \Big(9 + C r^2 (2 + C r^2)\Big) +
   7 K \Big(-3 + C r^2 (-10 + 9 C r^2)\Big)\bigg \}}{56 \pi (1 + C r^2)^3 (1 + 5 C r^2)^{7/5}},\\
p^{\text{eff}}&=&\frac{ C \bigg\{7 K (1 + 9 C r^2) - 16 (1 + 5 C r^2)^{2/5} \Big(-2 + C r^2 (7 + C r^2)\Big) \bigg\}}{56 \pi (1 + C r^2)^3 (1 + 5 C r^2)^{2/5}},
\end{eqnarray}

Using the expression of $p^{\text{eff}}$ and $\rho^{\text{eff}}$, from eqns. (\ref{r1}) and (\ref{r2}), we obtain the expression of matter density and pressure $\rho$ and $p$ in modified gravity as given by,
%%%%%% new
\begin{eqnarray}
\rho&=&\frac{ C }{7 (\chi + 2 \pi)(\chi + 4 \pi) (1 + C r^2)^3 (1 + 5 C r^2)^{7/5}}\Bigg[\pi \Bigg\{8 (1 + 5 C r^2)^{7/5} \Big(9 + C r^2 (2 +  C r^2)\Big) \nonumber\\&& +
   7 K \Big(-3 + C r^2 (-10 + 9 C r^2)\Big)\Bigg\} +
 \chi \Bigg\{ (1 + 5 C r^2)^{7/5} \Big(31 + C r^2 (-8 + C r^2)\Big)\nonumber\\&&
  +7 K \Big(-1 + C r^2 (-2 + 9 C r^2)\Big)\Bigg\}\Bigg],\label{p1}\\
      p&=&\frac{ C }{7 (\chi + 2 \pi)(\chi + 4 \pi) (1 + C r^2)^3 (1 + 5 C r^2)^{7/5}}\Bigg[7 K \Bigg\{\pi + 2 C (2 \chi + 7 \pi) r^2 + 9 C^2 (2 \chi + 5 \pi) r^4\Bigg\} \nonumber\\&& -(1 + 5 C r^2)^{7/5} \Bigg\{16 \pi \Big(-2 + C r^2 (7 + C r^2)\Big)+\chi \Big(-21 + 5 C r^2 (8 + C r^2)\Big)\Bigg\}\Bigg]. \label{p2}
\end{eqnarray}
\begin{figure}[htbp]
    \centering
        \includegraphics[scale=.55]{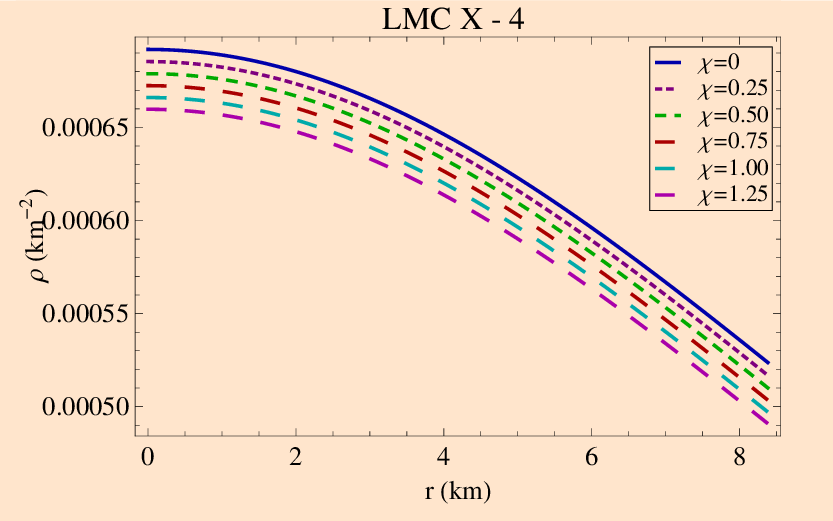}
        \includegraphics[scale=.55]{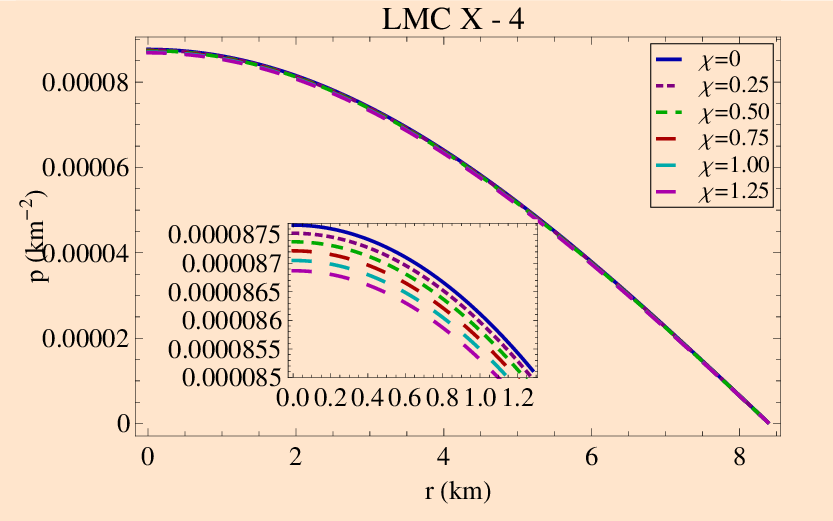}
        \caption{Profiles of (left) Matter density and (right) isotropic pressure with respect to radius. \label{pp}}
\end{figure}
The graphical representations of matter density and isotropic pressure are shown in Fig.~\ref{pp} for different values of $\chi$. We can verify from the figure that the two physical variables are maximum at the origin and decrease monotonically radially outward to reach their minimum values at the surface, proving the physical availability of our predicted stellar model. Interestingly, pressure drops to zero at the boundary of the star, though density does not. From these figures we can check that the energy density and pressure at the origin are positive and regular, indicating that our framework is free from physical and mathematical singularities.

\section{Exterior line element and junction conditions}\label{sec4}
At the boundary $r=R$ of the star, our interior spacetime will be matched to the Schwarzschild exterior solution at the boundary i.e., at $r=R$(Radius of the star). Corresponding to the interior spacetime,
 \begin{eqnarray}
ds_{-}^2 & = & A (1 + C r^2)^4 dt^2+ \Bigg[\frac{C K r^2}{(1 + C r^2)^2 (1 + 5 C r^2)^{2/5}} + \frac{7 - 10 C r^2 - C^2 r^4}{7 (1 + C r^2)^2}\Bigg]^{-1} dr^2+r^2(d\theta^2+\sin^2 \theta d\phi^2),
\end{eqnarray}
the exterior line element is described by,
\begin{eqnarray}
ds_+^{2}&=&-\left(1-\frac{2M}{r}\right)dt^{2}+\left(1-\frac{2M}{r}\right)^{-1}dr^{2}+r^{2}\left(d\theta^{2}+\sin^{2}\theta d\phi^{2}\right),
\end{eqnarray}
where `M' is the total mass within the boundary of the compact star.\\
Now following the continuity of the metric potentials at the boundary surface of the strange star $r = R$ :
$$g_{rr}^+=g_{rr}^-,\, ~~\text{and} ~~~~g_{tt}^+=g_{tt}^-,$$
where ($-$) and ($+$) signs represent interior and exterior spacetime, respectively. The above two relationships imply,
\begin{eqnarray}
\left(1-\frac{2M}{R}\right)^{-1}&=&\Bigg[\frac{C K R^2}{(1 + C R^2)^2 (1 + 5 C R^2)^{2/5}} + \frac{7 - 10 C R^2 - C^2 R^4}{7 (1 + C R^2)^2}\Bigg]^{-1},\label{o1}\\
1-\frac{2M}{R}&=& A (1 + C R^2)^4,\label{o2}
\end{eqnarray}
We also require that the isotropic pressure $`p'$ vanishes at the boundary of the star `R' i.e., $p(r=R)=0$, which implies that:
\begin{eqnarray}
7 K \Big(\pi + 2 C (2 \chi + 7 \pi) R^2 +
   9 C^2 (2 \chi + 5 \pi) R^4\Big) &=& (1 + 5 C R^2)^{7/5} \Big(16 \pi \big(-2 + C R^2 (7 + C R^2)\big) \nonumber\\&& +
   \chi \big(-21 + 5 C R^2 (8 + C R^2)\big)\Big) .\label{o3}
    \end{eqnarray}
 We obtain the expressions for $A$ and $K$ as function of $C$ by solving the equations (\ref{o1})-(\ref{o3}) simultaneously,
  \begin{eqnarray*}
 A &=& \frac{-2 M + R}{R (1 + C R^2)^4},\\
       K&=& \frac{(1 + 5 C R^2)^{7/5} \Bigg\{16 \pi \Big(-2 + C R^2 (7 + C R^2)\Big) +
   \chi \Big(-21 + 5 C R^2 (8 + C R^2)\Big)\Bigg\}} {7 \Bigg\{\pi + 2 C (2 \chi + 7 \pi) R^2 +
   9 C^2 (2 \chi + 5 \pi) R^4\Bigg\}}
      \end{eqnarray*}
The approximated mass and radius of the compact star LMC X-4 are used to determine the constant values of $A$ and $K$ which are presented in
Table~\ref{tb12} for various values of $\chi$ by choosing a preferred value of $C$.  It is intriguing to see from the table that numerical values of $A$ are constant because they are independent of $\chi$. On the contrary, the values of $K$ decrease with increasing values of $\chi$.

%\begin{figure}[htbp]
%    \centering
%        \includegraphics[scale=.55]{metric1.eps}
%        \includegraphics[scale=.55]{metric2.eps}
        %\includegraphics[scale=.6]{mass.eps}
%       \caption{The metric coefficients $e^{\lambda}$ and $e^{\nu}$ are plotted against radius for different values of the coupling constant $\beta$ mentioned in the figure. The interior spacetime is matched to the exterior spacetime at the boundary. \label{metric}}
%\end{figure}

\begin{table*}[t]
\centering
\caption{The numerically computed values of the constants $A$ and $K$ for the compact star LMC X-4 for different values of coupling parameter $\chi$(Taking $C=0.001 ~km^{-2}$).}
\label{tb12}
\begin{tabular}{@{}cccccccccccccccc@{}}
\hline
Star  & Estimated &Estimated & $\chi$ & $A$& $K$   \\
&Mass ($M_{\odot}$)& Radius &&&  \\
\hline
LMC X-4  \cite{Rawls:2011jw}& $1.04$&$8.4$& 0.00& 0.483244&$-2.36808$\\
&&& $0.25$&0.483244 &$-2.47737$                                    \\
&&& $0.50$ &0.483244  &$-2.58381$\\
&&& $0.75$ &0.483244  &$-2.68751$\\
&&& $1.00$ &0.483244 &$-2.78857$\\
&&& $1.25$ &0.483244 &$-2.88710$\\
\hline
\end{tabular}
\end{table*}

\section{Physical attributes}\label{sec5}

In this section, we will investigate the modified TOV equation, energy conditions, the status of the sound speed within the stellar system, effective compactness, surface redshift, gravitational redshift, adiabatic index, mass-radius relationship, and so on in the context of $f(R,\,T)$ theory for various values of the coupling parameter $\chi$.

\subsection{Pressure and density behavior of the compact object}
In order to verify the nonsingularity nature of the pressure and energy density, we evaluate the central pressure and central density as,
\begin{eqnarray}
\rho_c &=& \frac{C \Big\{ \chi (31 - 7 K) + 3\pi (24 - 7 K) \Big\}}{7 (\chi + 2 \pi) (\chi + 4 \pi)},\\
 p_c&=& \frac{C \Big\{ 21\chi + \pi (32 + 7 K) \Big\}}{7 (\chi + 2 \pi) (\chi + 4 \pi)}.
\end{eqnarray}
Hence both $\rho_c$ and $p_c$ are finite in nature.\\

Now we want to find a range for the coupling parameter $\chi$.\\
\begin{center}
  Case I  
\end{center}
We first assume that $\chi+2\pi >0$,\\
Now $p_c>0 \Rightarrow \chi>-\frac{\pi}{21}(32+7K)$\\
So we further obtain,
\begin{eqnarray}\label{x1}
\chi>max\left\{-2\pi, \,-\frac{\pi}{21}(32+7K)\right\},
\end{eqnarray}
Again $\rho_c-p_c>0 \Rightarrow (\chi+4\pi)(10-7K)>0$, which gives $10-7K>0$.\\
Now $\rho_c>0$ implies $\chi(31-7K)>-3\pi(24-7K)$,\\
Since, $10-7K>0$, the above inequality gives, $\chi>-3\frac{\pi(24-7K)}{31-7K}$.\\
By using the range of $K<10/7$, the range of $\chi$ given above further modified as, 
\begin{eqnarray}\label{x2}
\chi>-3\pi,
\end{eqnarray}
Combining (\ref{x1}) and (\ref{x2}), we get,
 \begin{eqnarray}
\chi>max\left\{-2\pi,\, -\frac{\pi}{21}(32+7K)\right\},
\end{eqnarray}
Now using $10>7K$, we get $-\frac{\pi}{21}(32+7K)>-2\pi$.\\
So finally we get a refined range for $\chi$ as, $\chi>-\frac{\pi}{21}(32+7K)$.\\

\begin{center}
    Case II
\end{center}
We assume $\chi+4\pi<0$, which gives, $\chi+2\pi<0$.\\
Now $\rho_c-p_c>0$ implies, $(\chi+4\pi)(10-7K)>0$ that gives, $10-7K<0$.\\
Now $p_c>0$ gives, $\chi>-\frac{\pi}{21}(32+7K)$.\\
So, under the condition $K>10/7$, we get a range for $\chi$ as,
\[-\frac{\pi}{21}(32+7K)<\chi<-2\pi,\] in this case.\\
\begin{center}
   Case III 
\end{center}
In this case we assume that $-4\pi <\chi<-2\pi$.\\
Now, $p_c>0$ gives, $\chi<-\frac{\pi}{21}(32+7K)$.\\
$\rho_c-p_c>0$ implies, $(\chi+4\pi)(10-7K)>0$ that gives, $10-7K>0$.\\
So under the circumstances $K<10/7$, we get a range for $\chi$ as,
\[-4\pi <\chi<min\left\{-2\pi,- \frac{\pi}{21}(32+7K)\right\}.\]

The pressure and density gradient can be obtained by taking the derivative of the expressions of $\rho$ and $p$ given in Eqns.(\ref{p1})-(\ref{p2}), which gives,
\begin{eqnarray*}
\rho'&=& \frac{ 2C^2 r }{7 (\chi + 2 \pi)(\chi + 4 \pi) (1 + C r^2)^4 (1 + 5 C r^2)^{12/5}}\Bigg[\chi \Bigg\{- (1 + 5 C r^2)^{12/5} \Big(101 + C r^2 (-18 +  C r^2)\Big) \nonumber\\&& +
   28 K \Bigg(2 + C r^2 \Big(12 + C r^2(13 - 27C r^2 )\Big)\Bigg)\Bigg\} -
 4 \pi \Bigg\{2 (1 + 5 C r^2)^{12/5} \Big(25 + C r^2 (2 +  C r^2)\Big)\nonumber\\&&
  +7 K \Big(-5 + C r^2 \big(-31 + C r^2 (-47 + 27 C r^2)\big)\Big)\Bigg\}\Bigg] ,\\
p'&=& \frac{ 2C^2 r }{7 (\chi + 2 \pi)(\chi + 4 \pi) (1 + C r^2)^4 (1 + 5 C r^2)^{12/5}}\Bigg[4 \pi (1 + 5 C r^2) \Bigg\{4 (-1 + C r^2) (13 + C r^2) (1 + 5 C r^2)^{7/5} \nonumber\\&& -
   7 K \Big(-1 + C r^2 (2 + 27 C r^2)\Big)\Bigg\} +
 \chi \Bigg\{ (1 + 5 C r^2)^{12/5} \Big(-103 + 5C r^2 (14 +  C r^2)\Big)\nonumber\\&&
  -28 K \Big(-1 + C r^2 \big(-5 + 8C r^2 +54 C^2 r^4\big)\Big)\Bigg\}\Bigg].
\end{eqnarray*}

\begin{figure}[htbp]
    \centering
        \includegraphics[scale=.55]{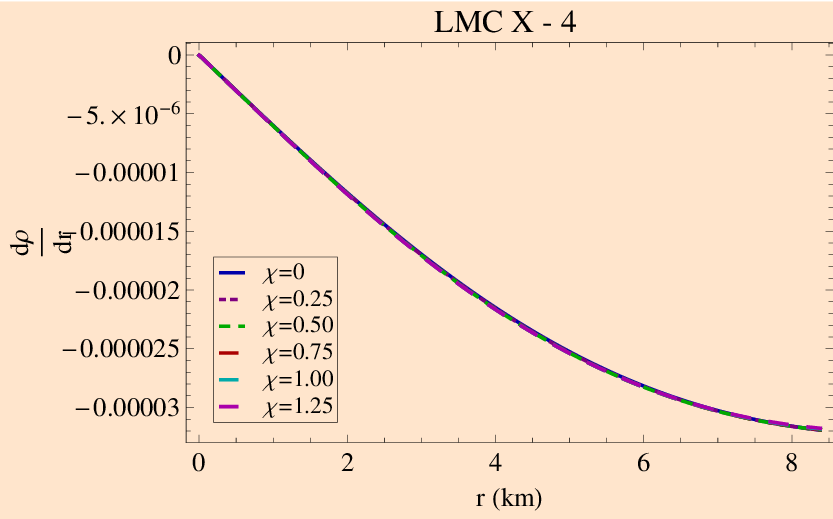}
        \includegraphics[scale=.55]{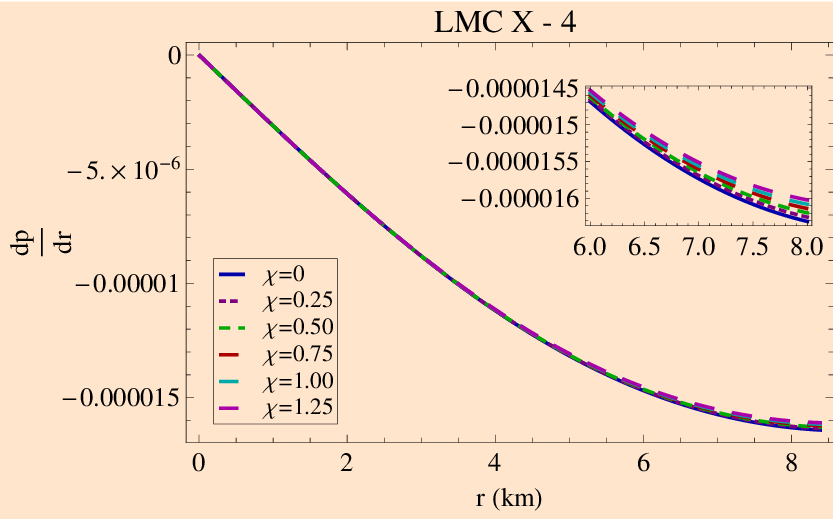}
       \caption{The pressure and density gradients with respect to `r'.}\label{grad5}
\end{figure}
\begin{figure}[htbp]
    \centering
        \includegraphics[scale=.55]{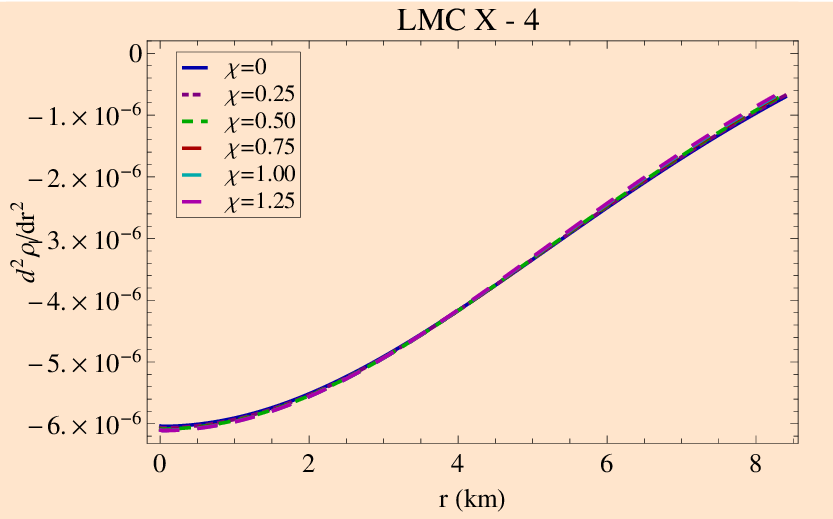}
        \includegraphics[scale=.55]{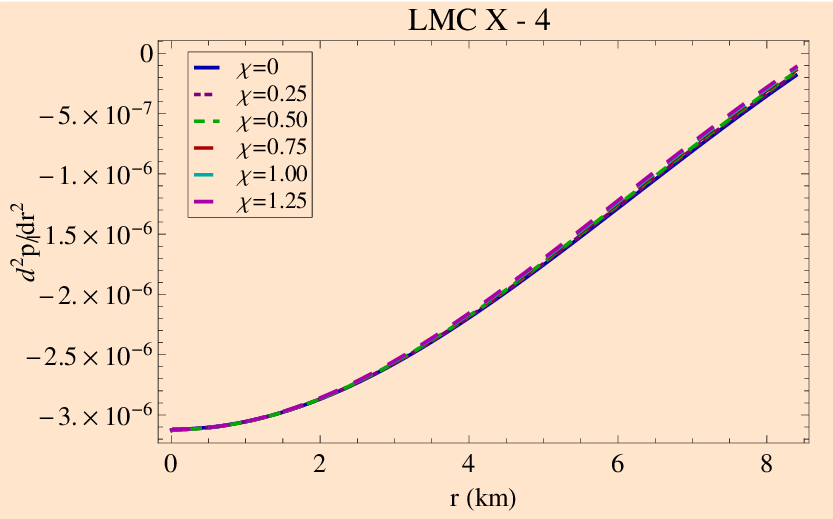}
       \caption{$\frac{d^2\rho}{dr^2}$ and $\frac{d^2 p}{dr^2}$ are shown with respect to `r'.}\label{rho2}
\end{figure}

The behavior of density and pressure gradient are shown in Fig.~\ref{grad5} for different values of $\chi$. From the figures, one can check that $\rho',\,p'<0$ in the interior of the stellar model and $\rho'(0)=0=p'(0)$. Moreover at the center of the star $\rho''(0),\,p''(0)<0$ as shown in Fig.~\ref{rho2}.\par
%\begin{figure}[htbp]
%    \centering
%        \includegraphics[scale=.55]{ec1.eps}
%        \includegraphics[scale=.55]{ec2.eps}
%        \includegraphics[scale=.55]{ec3.eps}
 %      \caption{All the energy conditions are plotted against the radius for different values of the coupling constant mentioned in the figure.\label{ec}}
%\end{figure}

 \subsection{Energy Conditions}
 The term "energy conditions" refers to a set of physical properties that can be used to investigate the presence of ordinary and exotic matter inside a star formation. The validity of the second law of black hole thermodynamics and the Hawking-Penrose singularity theorems can be easily tested using the energy conditions \cite{Hawking:1973uf}. These conditions of energy are referred to as null, weak, strong, and dominant energy conditions, symbolized respectively by NEC, WEC, SEC, and DEC. We tested all of the energy conditions in our research. If the following inequalities hold then all energy conditions will be satisfied:
    \begin{eqnarray*}
 \text{WEC}:~\rho+p\geq 0,~ \rho \geq 0,\, \text{NEC}:~\rho+p\geq 0,\, \text{SEC}:~\rho+p \geq 0, \rho+ 3p \geq 0,\,\text{DEC}:~\rho-p\geq 0,~ \rho \geq 0.
\end{eqnarray*}
We will need the following expressions to check the aforementioned energy conditions:
\begin{eqnarray}
\rho+p&=&\frac{C \Bigg\{4 (1 - C r^2) (13 + C r^2) (1 + 5 C r^2)^{7/5} +
   7 K \Big(-1 + C r^2 (2 + 27 C r^2)\Big)\Bigg\}}{7 (\chi + 4 \pi) (1 + C r^2)^3 (1 + 5 C r^2)^{7/5}},\\
\rho+3p&=&\frac{ C }{7 (\chi + 2 \pi)(\chi + 4 \pi) (1 + C r^2)^3 (1 + 5 C r^2)^{7/5}}\Bigg[\chi \Big(-2 (1 + 5 C r^2)^{7/5} \big(-47 + C r^2 (64 + 7 C r^2)\big) \nonumber\\&& +
   7 K \big(-1 + C r^2 (10 + 63 C r^2)\big)\Big) -
 8 \pi \Big\{-21 (1 + 5 C r^2)^{2/5} + C r^2 \Big(-14 K (2 + 9 C r^2)\nonumber\\&&
  +5 (1 + 5 C r^2)^{2/5} \big(-13 + C r^2 (41 + 5 C r^2)\big)\Big)\Big\}\Bigg],\\
   \rho-p&=&\frac{ C (1 + 3 C r^2) \Big(-7 K (1 + 3 C r^2) + 2 (5 + C r^2) (1 + 5 C r^2)^{7/5}\Big) }{7 (\chi + 2 \pi) (1 + C r^2)^3 (1 + 5 C r^2)^{7/5}},
\end{eqnarray}

 \begin{figure}[htbp]
    \centering
        \includegraphics[scale=.39]{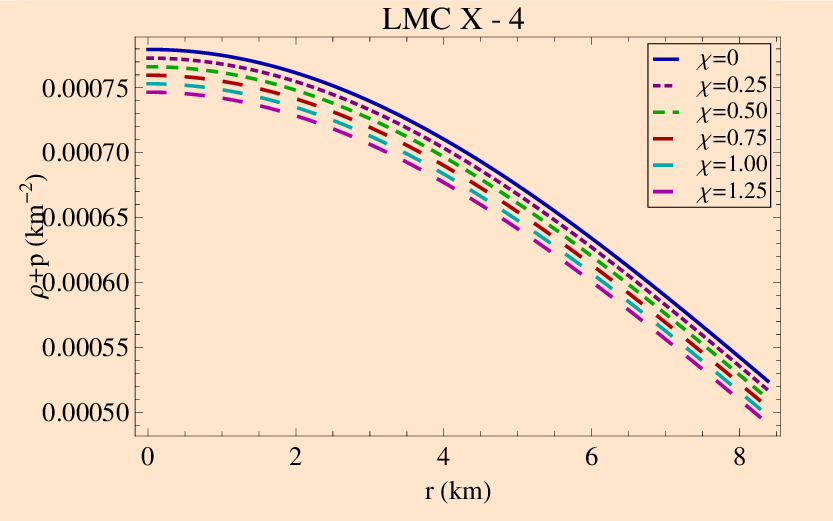}
        \includegraphics[scale=.39]{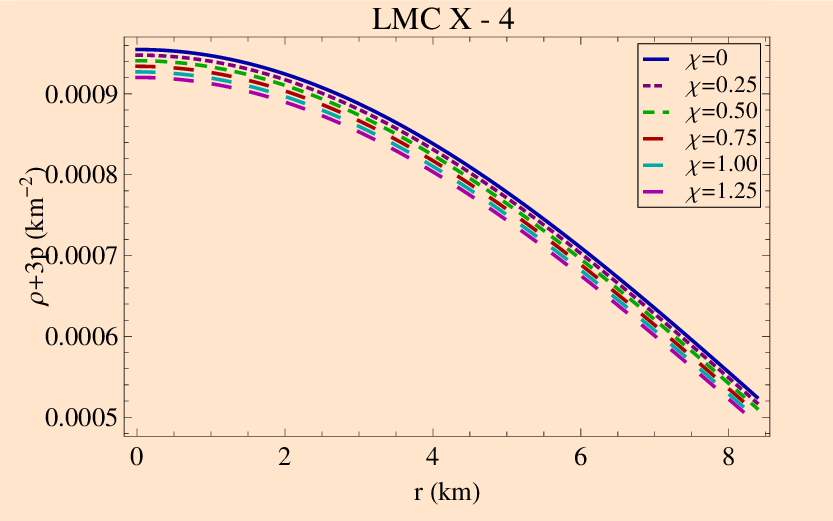}
        \includegraphics[scale=.39]{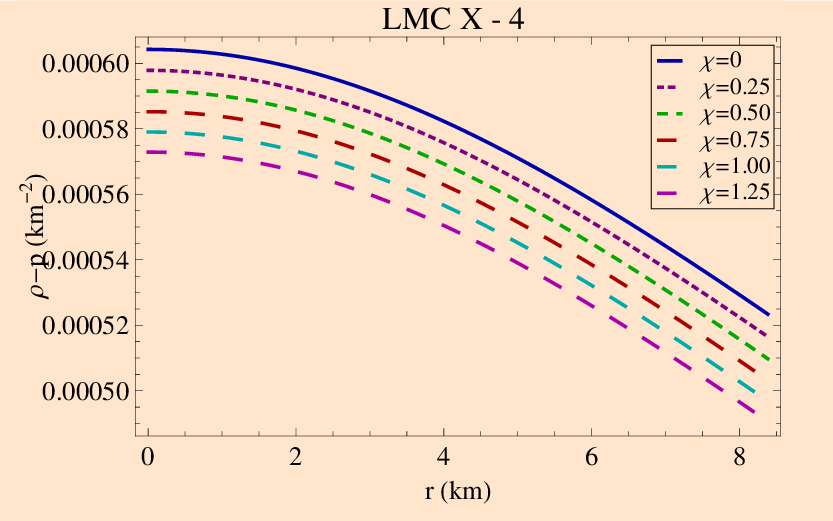}
       \caption{The energy conditions are plotted against `r'.}\label{ener1}
\end{figure}

As illustrated graphically in Fig.~\ref{ener1}, all energy conditions for our model have been met.

\subsection{Causality condition via sound velocity behavior}

The stability criterion for a physically viable stellar compact object is now being investigated using graphical analysis utilizing numerical values for several unknown constants. This criterion is demonstrated using the causality condition. For this, the situation of sound velocity inside the stellar model is reviewed in this section. The square of the velocity of sound $V^2$ in the entire region of the fluid sphere must lie in the desired range $(0,1)$, which is termed as the causality condition.
\begin{eqnarray}
V^2&=&\frac{h_1(r)}{h_2(r)},
\end{eqnarray}
where,
\begin{eqnarray*}
h_1(r) &=&  -4 \pi (1 + 5 C r^2) \Big \{4 (-1 + C r^2) (13 + C r^2) (1 + 5 C r^2)^{7/5} -
    7 K \big(-1 + C r^2 (2 + 27 C r^2)\big)\Big \} \\&& + \chi \Big \{(1 + 5 C r^2)^{12/5}  \big(103 - 5 C r^2 (14 + C r^2)\big) + 28 K \big(-1 + C r^2 (-5 + 8 C r^2 + 54 C^2 r^4)\big)\Big \}  ,\\
h_2(r) &=& \chi \Big \{(1 + 5 C r^2)^{12/5} (101 + C r^2 (-18 + C r^2)) +
    28 K (-1 + C r^2) (2 + C r^2 (14 + 27 C r^2))\Big \} \\&& +  4 \pi \Big\{2 (1 + 5 C r^2)^{12/5}   \big(25 + C r^2 (2 + C r^2)\big) +   7 K \Big(-5 + C r^2 \big(-31 + C r^2 (-47 + 27 C r^2)\big)\Big)\Big\}
\end{eqnarray*}
\begin{figure}[htbp]
    \centering
        \includegraphics[scale=.55]{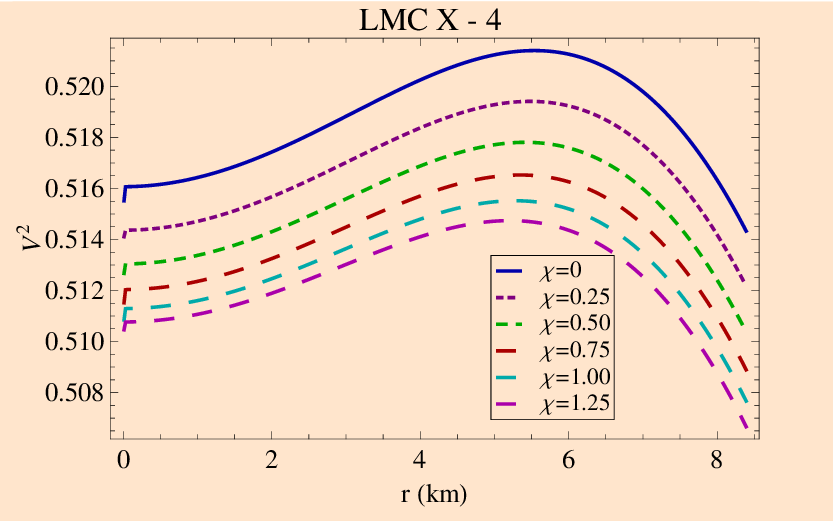}
        \includegraphics[scale=.55]{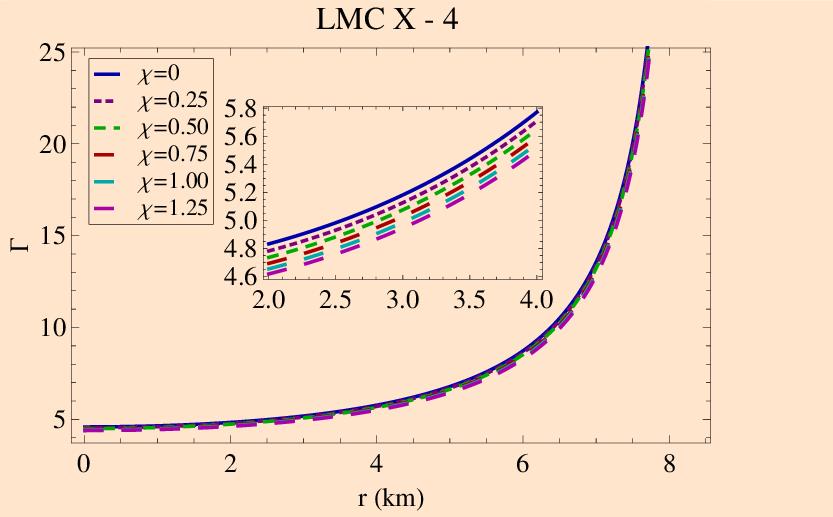}
       \caption{(left) The square of the sound velocity and (right) relativistic adiabatic index are plotted against the radius inside the stellar interior.\label{sv}}
\end{figure}
The graphical aspect of the causality condition for our chosen compact object LMC X-4 is investigated in Fig.~\ref{sv} for different values of $\chi$, from which it is evident that across the fluid sphere, the square of the sound velocity is within our expected range.

\subsection{Status of relativistic adiabatic index inside $f(R,\,T)$ gravity}

The relativistic adiabatic index, which also describes the stability of both relativistic and non-relativistic compact objects, can be used to describe the stiffness of the equation of state for a given energy density. The idea of dynamical stability against infinitesimal radial adiabatic disturbance of the stellar system was first proposed by Chandrasekhar \cite{Chandrasekhar:1964zz}. Later, this stability condition has been effectively illustrated in various astrophysical investigations \cite{bardeen1966catalogue,heintzmann1975neutron,hillebrandt1976anisotropic,knutsen1988stability,mak2013isotropic} for both isotropic and anisotropic stellar objects. According to their calculations, the adiabatic index of a dynamically stable stellar object must be greater than $4/3$ at all points within the compact object. The relativistic adiabatic index ($\Gamma$) is expressed as,
\begin{eqnarray}
\Gamma&=&\frac{\rho+p}{p}V^2,\nonumber\\&=&\frac{h_3(r)}{h_4(r)}
\end{eqnarray}
where,
\begin{eqnarray*}
h_3(r) &=& -\Bigg[(\chi + 2 \pi) \Bigg\{4 (-1 + C r^2) (13 + C r^2) (1 + 5 C r^2)^{7/5} -
     7 K \big(-1 + C r^2 (2 + 27 C r^2)\big)\Bigg\} \Bigg\{4 \pi (1 + 5 C r^2) \times \\&& \Bigg(4 (-1 + C r^2) (13 + C r^2) (1 + 5 C r^2)^{7/5} -7 K \Big(-1 + Cr^2 (2 + 27 C r^2)\Big)\Bigg) + \chi \Bigg((1 + 5 C r^2)^{12/5} \times \\&& \Big(-103 + 5 C r^2 (14 + C r^2)\Big) - 28 K \Big(-1 + C r^2 \big(-5 + 8 C r^2 + 54 C^2 r^4)\big)\Big)\Bigg\}\Bigg],\\
h_4(r) &=& \Bigg[\Bigg\{- 7 K \Bigg(\pi + 2 C (2 \chi + 7 \pi) r^2 +
    9 C^2 (2 \chi + 5 \pi) r^4\Bigg) + (1 + 5 C r^2)^{7/5} \Bigg(16 \pi \Big(-2 + C r^2 (7 + C r^2)\Big) \\&& +
     \chi \Big(-21   + 5 C r^2 (8 + C r^2)\Big)\Bigg)\Bigg\}  \Bigg\{\chi \Bigg((1 + 5 C r^2)^{12/5} \Big(101 + C r^2 (-18 + C r^2)\Big) + 28 K (-1 + C r^2) \times \\&& \Big(2 + C r^2 (14 + 27 C r^2)\Big)\Bigg)   +4 \pi \Bigg(2 (1 + 5 C r^2)^{12/5} \Big(25 + C r^2 (2 + C r^2)\Big) + 7 K \Big(-5 + C r^2 \big(-31 \\&& + C r^2 (-47 + 27 C r^2)\big)\Big)\Bigg)\Bigg\}\Bigg],\\.
\end{eqnarray*}
In Fig.~\ref{sv}, we see the behavior of the adiabatic index $\Gamma$ which depicts that the value of the adiabatic index is much greater than $4/3$ everywhere in the interior of the stellar object.  Thus, we can conclude the stability of our proposed model.
\subsection{Dynamical equilibrium via modified TOV equation within $f(R,\,T)$ gravity}
The hydrostatic equilibrium equation is a key feature of our presented physically realistic and stable compact stellar object. By using the generalized Tolman-Oppenheimer-Volkov (TOV) equation, we can check the equilibrium condition of our obtained model for the chosen stellar candidate under the combined behavior of different forces acting on our present system as:
\begin{eqnarray}\label{con1}
-\frac{\nu'}{2}(\rho+p)-\frac{dp}{dr}+\frac{\chi}{8\pi+2\chi}(p'-\rho')=0,
\end{eqnarray}

The above modified TOV equation (\ref{con1}) describes the equilibrium condition for our present isotropic stellar model in $f(R,\, T)$ gravity system. The Eqn. (\ref{con1}) can be split into three different forces, namely gravitational ($F_g$), hydrostatic ($F_h$), and the additional force quantity ($F_m$) due to the coupling between the geometry and the matter in modified gravity, where $F_g = -\frac{\nu'}{2}(\rho+p)$, $F_h = -\frac{dp}{dr}$ and $F_m = \frac{\chi}{8\pi+2\chi}(p'-\rho')$. If the sum of these three different forces is zero i.e. $F_g + F_h + F_m = 0$, then this equation implies that this $f(R,\,T)$ stellar model is in stable equilibrium. Also to be stable, the coupling parameter value should be $\chi \neq -4\pi$.  Here, the force $F_m$ in our present model describes the coupling between the geometry and the matter within the modified gravity. For the choice of $\chi=0$ our present model reduces to General Relativity (GR) and in that case, the TOV equation reduces to 
\begin{eqnarray}\label{tov}
-\frac{\nu'}{2}(\rho+p)-\frac{dp}{dr}=0,
\end{eqnarray}
which coincides with the results of Oppenheimer and Volkoff \cite{Oppenheimer:1939ne}.
\\
\begin{figure}[htbp]
    \centering
        \includegraphics[scale=.55]{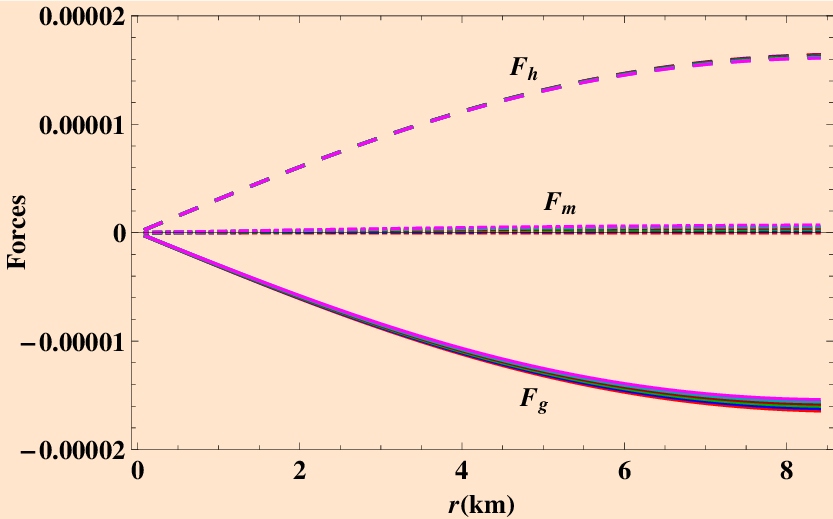}
       \caption{Variation of forces acting on the system plotted against the radius at the stellar interior. \label{tov1}}
\end{figure}

In our case,
\begin{eqnarray}
F_g&=& \frac{4 C^2 r \Big\{ 4 (-1 + C r^2) (13 + C r^2) (1 + 5 C r^2)^{7/5} -
   7 K \Big(-1 + C r^2 (2 + 27 C r^2)\Big)\Big\}}{7 (\chi + 4 \pi) (1 + C r^2)^4 (1 + 5 C r^2)^{7/5}},\\
F_h &=&\frac{1}{7 (\chi + 2 \pi) (\chi + 4 \pi) (1 + C r^2)^4 (1 + 5 C r^2)^{
 12/5}}\Bigg[ 2 C^2 r \Bigg\{-4 \pi (1 +
      5 C r^2) \Big(4 (-1 + C r^2) (13 + c r^2) (1 + 5 C r^2)^{7/5} \nonumber\\
     && -7 K \big(-1 + C r^2 (2 + 27 C r^2)\big)\Big) +
   \chi \Big((1 + 5 C r^2)^{12/5} \big(103 - 5 C r^2 (14 + C r^2)\big) +
      28 K \big(-1 + C r^2 (-5 + 8 C r^2 \nonumber\\
     && + 54 C^2 r^4)\big)\Big)\Bigg\} \Bigg],\\
   F_m &=& \frac{2 C^2 \chi r \Big\{(1 + 5 C r^2)^{12/5} \Big(-1 + C r^2 (26 + 3 C r^2)\Big) -
   14 K (1 + 3 C r^2) \Big(1 + C r^2 (4 + 9 C r^2)\Big)\Big\}}{7 (\chi + 2 \pi) (\chi + 4 \pi) (1 + C r^2)^4 (1 + 5 C r^2)^{
 12/5}}.
   \end{eqnarray}
      The profiles of all the forces involved in the hydrostatic equilibrium condition have been shown in Fig.~\ref{tov1}. It can be clearly observed from the figure that the gravitational force ($F_g$) is attractive in nature and is acting inwards. Whereas, hydrostatic force ($F_h$) is acting in an outward direction by which we can conclude that it is repulsive in nature. The force $F_h$ attains its maximum near the boundary. On the other hand, the coupling force $F_m$ has a very negligible effect to achieve stable equilibrium.

      Finally, we can conclude from the figure that the gravitational force counterbalances the combined action of hydrostatic force and coupling force inside the star to keep our present model in stable equilibrium.
      
\subsection{Equation of state} 

\begin{figure}[htbp]
    \centering
        \includegraphics[scale=.55]{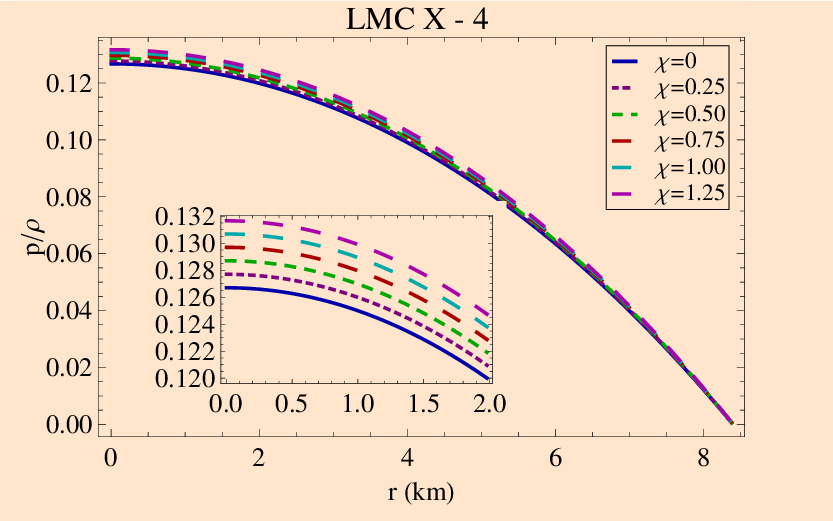}
        \includegraphics[scale=.55]{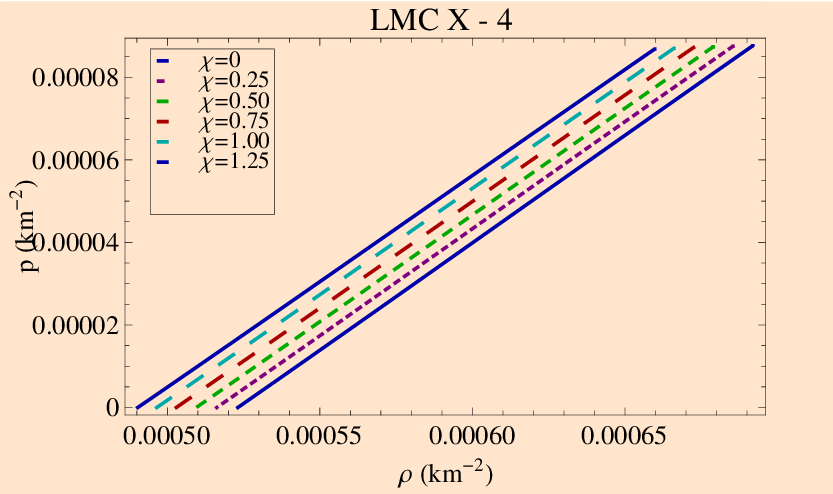}
       \caption{The pressure and density relation are shown at the stellar interior.}\label{eoss}
\end{figure}

The link between pressure and matter density is primarily described by the equation of state. Many academics have already used linear, quadratic, polytropic, and other forms of equations of state (EoS) to model the compact object. We did not use any particular equation of state to create the stellar model in our work. We used a graphical approach to depict the variation of pressure with respect to density in Fig.~\ref{eoss}. For varying values of $\chi$, the pressure-to-matter density ratio is also shown in Fig.~\ref{eoss}.

\subsection{Mass-Radius relationship}

The mass function $m(r)$ of our present model is determined by using energy density($\rho$) as,
\begin{eqnarray}
m(r)&=&4\pi\int_0^r \rho r^2 dr,\nonumber\\
&=&\frac{1}{5040 \sqrt{C} (\chi + 2 \pi) (\chi + 4 \pi) (1 + C r^2)}\pi \Bigg[90 \sqrt{C} (745 \chi + 584 \pi) r -
   6720 C^{3/2}
     K r^3 (\chi + 3 \pi) (1 + C r^2)\nonumber\\
     && \times F_1 (3/2; 3, 7/5; 5/2; -C r^2, -5 C r^2) -
   8064 C^{5/2}
     K r^5 (\chi + 5 \pi)  (1 + C r^2) F_1 \big(5/2; 3, 7/5; 7/2; -C r^2, -5 C r^2\big) \nonumber\\
     && + (1 +
      C r^2) \Bigg\{25920 C^{7/2}
        K r^7 (\chi + \pi)  F_1 \big(7/2; 3, 7/5; 9/2; -C r^2, -5 C r^2\big) +
      1600 C^{9/2} r^9 (\chi + 8 \pi) \nonumber\\
     && F_1 \big(9/2, 3, 1, 11/2, -C r^2, -5 C r^2\big) +
      9 \Bigg(5 (991 \chi + 2296 \pi) \tan^{-1}({\sqrt{C} r}) -
         \sqrt{5}~ (\chi + 8 \pi) \tan^{-1}({\sqrt{5} \sqrt{C} r}) \nonumber\\
     && -80r \sqrt{C} (155 \chi + 216 \pi)  _2F_1\big(1/2, 3, 3/2; -C r^2\big)\Bigg)\Bigg\}\Bigg]
 \end{eqnarray}
 The result of the integration is very complicated due to the presence of Appell hypergeometric function $F_1 (a;b_1,b_2;c;x,y)$ and hypergeometric function $_2F_1(a,b,c;z)$ in its expression. Appell hypergeometric function $F_1 (a;b_1,b_2;c;x,y)$ reduces to ordinary hypergeometric function $_2F_1(a,b,c;z)$  when $x=0$  or $y=0$. One can easily check that the mass function depends on the coupling parameter $\chi$. The effective mass function of the system can be calculated as :
 \begin{eqnarray}
 m^{\text{eff}}(r)&=&4\pi\int_0^r \rho^{\text{eff}} r^2 dr=\frac{C r^3}{14 (1 + C r^2)^2}\bigg\{24 + 8 C r^2 - \frac{7 K}{(1 + 5 C r^2)^{2/5}}\bigg\},
 \end{eqnarray}
The effective mass of a compact star is directly proportional to its radius, as verified by the behavior of the mass function in Fig.~\ref{m11}. Clearly, the mass function is regular at the core. From this graph, we can see that the maximum mass is achieved at the boundary of the star.

\subsection{Effective compactness factor, surface redshift, and gravitational redshift}

\begin{figure}[htbp]
    \centering
        \includegraphics[scale=.45]{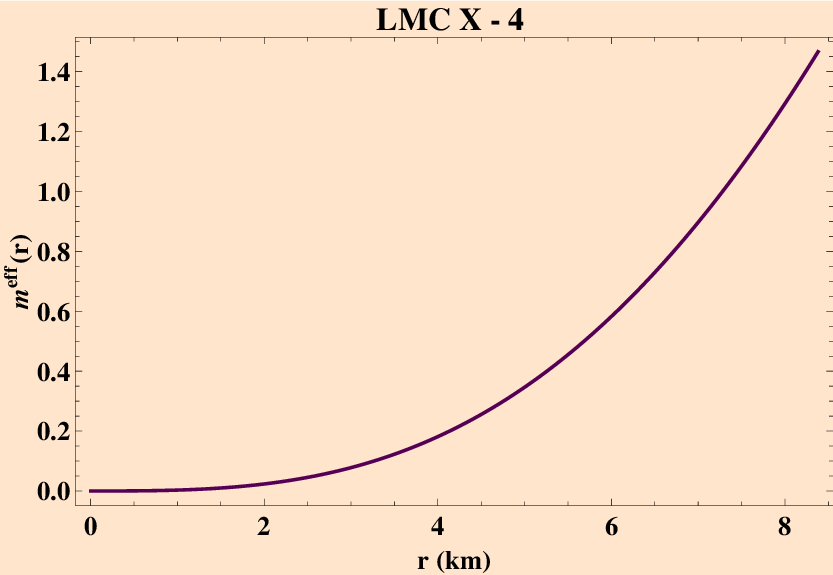}
        \includegraphics[scale=.45]{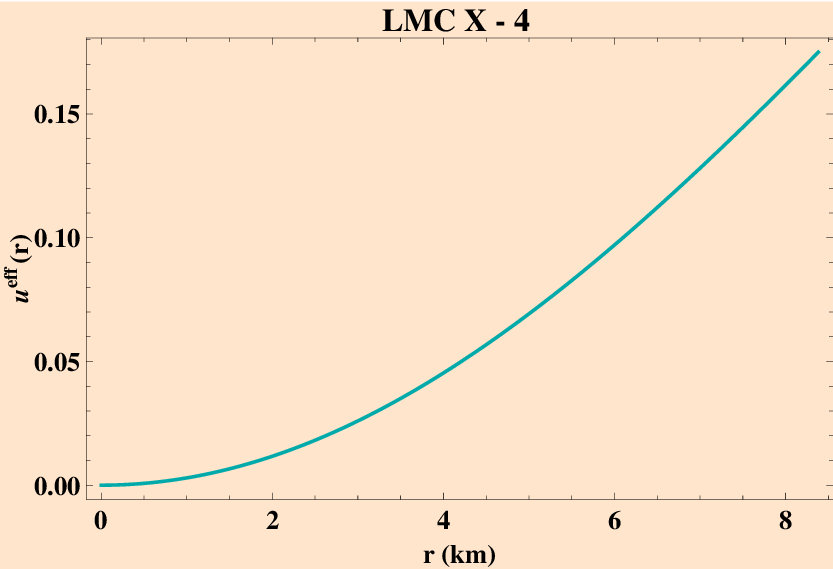}
        \includegraphics[scale=.45]{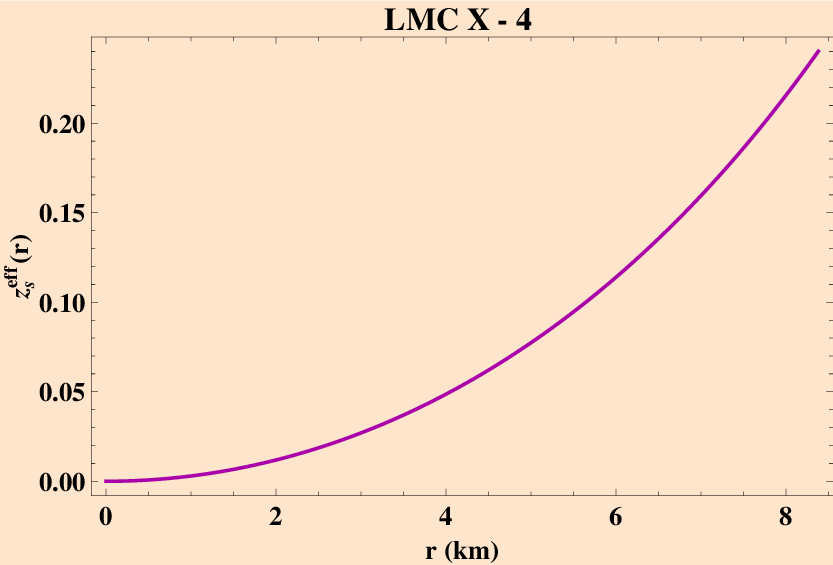}
        \includegraphics[scale=.47]{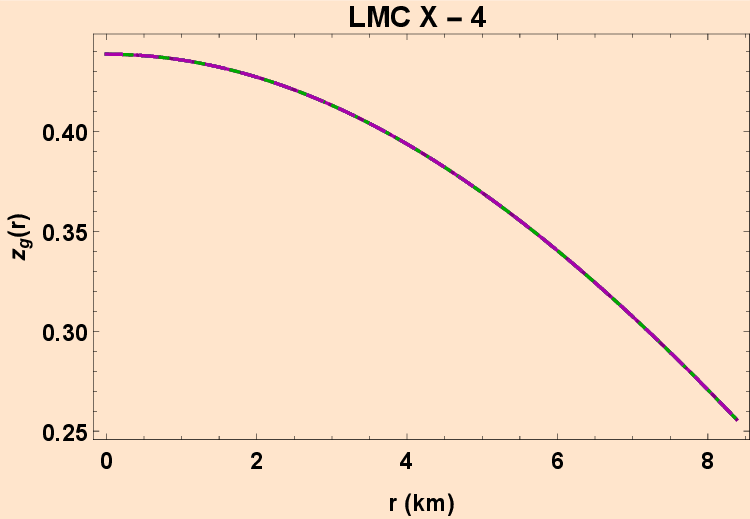}
       \caption{(upper left) Effective mass, (upper right) effective compactness, (lower left) effective surface redshift, and (lower right) gravitational redshift are plotted against the radius ($r$) inside the stellar interior.\label{m11}}
\end{figure}

The effective compactness factor $u^{\text{eff}}$ which classifies the compact objects into different categories as normal star ($u^{\text{eff}} \sim 10^{-5}$), white dwarfs ($u^{\text{eff}}~\sim 10^{-3}$), a neutron star ($10^{-1} <~u^{\text{eff}}~< 1/4$), ultra-compact star ($1/4<~u^{\text{eff}}<1/2$)
and black hole ($u^{\text{eff}} \sim 1/2$) can be expressed in terms of mass function as follows :
\begin{eqnarray}
u^{\text{eff}}=\frac{m^{\text{eff}}}{r}=\frac{C r^2}{14 (1 + C r^2)^2}\bigg\{24 + 8 C r^2 - \frac{7 K}{(1 + 5 C r^2)^{2/5}}\bigg\}.
\end{eqnarray}
Furthermore, the following relation can be used to calculate surface redshift($z_s^{\text{eff}}$):
\begin{eqnarray}
z_s^{\text{eff}}&=&\frac{1}{\sqrt{1-2u^{\text{eff}}}}-1,
\end{eqnarray}
The behavior of effective compactness and surface redshift inside the stellar object is shown in Fig.~\ref{m11}. From this plot, we can check that the maximum values of effective compactness and surface redshift are achieved at the boundary of the star.

Whereas the gravitational redshift (also known as interior redshift) denoted by $z_g(r)$ within a static line element can be calculated by the temporal component of the gravitational metric potentials as,
$z_g(r)=\sqrt{e^{-\nu(r)}}-1$. We have displayed the variation of $z_g(r)$ in Fig.~(\ref{m11}). The behavior of gravitational redshift also depicts the nature of interior matter density. It is evident from the plot that $z_g(r)$ is maximum at the core and gradually decreases with the increasing radius to attain its minimum value at the stellar boundary.

From Figure~(\ref{m11}), it is very clear that gravitational and surface redshift trends are totally opposite to one another throughout the fluid sphere. The reason behind this is: if a photon emits from the center to reach the surface, it has to travel a longer route and through a significantly denser region (the core). This causes additional dispersion, which causes energy loss.
On the other side, a photon that emerges from a location near the surface will take a shorter route through a less dense zone, resulting in reduced dispersion and energy loss. The interior redshift is, therefore, greatest near the center and lowest at the surface. However, the surface redshift is influenced by the total mass and radius, or surface gravity. As mass increases the radius will also increase slightly which will result in more surface gravity and more surface redshift.

\begin{table*}[t]
\centering
\caption{The numerically computed values of central density($\rho_c$), surface density($\rho_s$), central pressure($p_c$) and central relativistic adiabatic index $\Gamma(r=0)$ for the compact star LMC X-4 for different values of coupling parameter $\chi$ (Taking $C=0.001 ~km^{-2}$).}
\label{tb1}
\begin{tabular}{@{}cccccccccccccccc@{}}
\hline
$\chi$& $\rho_c (\rm{gm.cm^{-3}})$ & $\rho_s (\rm{gm.cm^{-3}})$ & $p_c (\rm{dyne.cm^{-2}})$ & $\Gamma(r=0)$\\
\hline
 0.00& $9.33632 \times 10^{14}$& $7.05853 \times 10^{14}$ & $1.06464 \times 10^{35}$&4.58915\\
 0.25& $9.24811 \times 10^{14}$ & $6.96534 \times 10^{14}$& $1.06291 \times 10^{35}$&4.54218\\
 0.50& $9.1605 \times 10^{14}$ & $6.87458 \times 10^{14}$ & $1.06109 \times 10^{35}$&4.49926\\
 0.75& $9.0736 \times 10^{14}$ & $6.78616 \times 10^{14}$ & $1.05917 \times 10^{35}$&4.45986\\
 1.00& $8.9875 \times 10^{14}$ & $6.69999 \times 10^{14}$ & $1.05712 \times 10^{35}$&4.42355\\
 1.25& $8.90228 \times 10^{14}$& $6.61597 \times 10^{14}$ & $1.05493 \times 10^{35}$&4.38996\\
\hline
\end{tabular}
\end{table*}

\section{Discussion}
In this paper, we offer a model for the compact star LMC X-4 and a fresh approach to Einstein field equations in $f(R, T)$ gravity. LMC X-4, an accretion-powered HMXB pulsar, has an orbital period of 1.41 days and a super orbital X-ray cycle. It is approximately 163,000 light years away, in the Large Magellanic Cloud (LMC). A pulsar, which is a highly magnetized neutron star radiating X-rays, and a companion star make up this two-star system LMC X-4. Heemskerk \& van Paradijs \cite{heemskerk1989analysis} concluded that the object included a twisted precessing accretion disc after undertaking a thorough investigation of the long-term oscillations in the X-ray flux of LMC X-4. Here, we have examined several physical behaviors of the suggested strange star in Durgapal-IV spacetime, both analytically and graphically. The key findings are highlighted through the graphical
representations and tables which are given as follows:
\begin{itemize}
  \item As can be seen from Fig.~\ref{metric}, the metric potentials are regular and devoid of any singularities. From the figure, it is clear that the $e^{\nu}$ profile does not depend on the value of $\chi$, although $e^{\lambda}$ assumes higher values for larger values of $\chi$.
  \item The physical variables pressure and density are well-defined inside the compact star structures. These parameters are represented graphically in Fig.~\ref{pp}. According to a thorough analysis of these figures, pressure, and density, both have positive definite values inside of the stellar configurations and are monotonically decreasing functions of the radial coordinate `r'. The pressure vanishes at the boundary of the stellar configuration. The statistics demonstrate that pressure and density decrease as the value of $\chi$ rises. According to our observations, the star gets less compact as the value of $\chi$ increases.
 \item In Table~\ref{tb12}, we represent the numerically computed values of the constants A and K for the compact star LMC X-4 for different values of coupling parameter $\chi$ by taking $C = 0.001 ~\rm{km^{-2}}$. From this table, we notice that the value of the constant K remains negative throughout the range of $\chi$ and A remains fixed as it does not depend on the value of $\chi$.
      \item The pressure and density gradients are shown graphically in Fig.~\ref{grad5}. At the stellar core, pressure, and density gradients no longer exist. Also, $\frac{d^2p}{dr}$ and $\frac{d^2\rho}{dr}$ take negative values (shown in Fig.~\ref{rho2}) at the center, which confirms that these parameters are monotonically decreasing inside the compact star.
          \item Fig.~\ref{ener1} shows that the compact star model obeys all the energy conditions inside the stellar interior for different values within $0\leq \chi \leq 1.5$.
          \item The causality condition is one of the most significant features of the stellar configuration. In Fig.~\ref{sv} the square of the sound velocity ($V^2$) is depicted graphically. The velocity of sound must be less than the speed of light for a stable stellar object and the values of $V^2$ must lie between the interval [0, 1] which are satisfied by our model. As a result, the model is realistic and the causality criterion is satisfied.
          \item In Fig.~\ref{sv} the plot of the relativistic adiabatic index ($\Gamma$) is shown for different values of $\chi$ in  $f(R,\,T)$ modified gravity. From figure, we see that $\Gamma$ is a monotonic increasing function of radial coordinate 'r' and takes values greater than $4/3$ everywhere inside the fluid sphere for every value of $\chi$. This guarantee the stability of our current model. The star becomes more stable as the value of $\chi$ increases (illustrated in the figure).
          \item Fig.~\ref{tov1} confirms that our proposed star model maintains hydrostatic equilibrium. As a result, our suggested model is stable when subjected to the effects of the active forces.
          \item In Fig.~\ref{eoss} we have displayed the variation of pressure with respect to density and also the pressure-density ratio as a function of 'r' for different values of $\chi$. We can observe from this figure that our fluid model is satisfying the Zeldovich condition i.e. pressure-density ratio is lying between 0 and 1.
          \item Table~\ref{tb1} represents the values of different physical parameters computed numerically for different values of the coupling parameter $\chi$ by taking $C = 0.001 km^{-2}$. In this table, we show the different values of surface density($\rho_s$), central density($\rho_c$), central pressure($p_c$), central relativistic adiabatic index $\Gamma(r=0)$ for the compact star LMC X-4 corresponding to the previously stated values of $\chi$. Surface density and central pressure are on a downward trend as $\chi$ values increase.
           \item Also, the effective mass function and effective compactness factor are well-behaved inside compact stellar formations, as seen in Figs.~\ref{m11} for various values of $\chi$. For all values of $\chi$, the effective surface redshift is maximum at the surface and increases monotonically as one proceeds from the center to the boundary, as shown in Fig.~\ref{m11}. From the figure, it is clear that the effective compactness factor is less than $4/9$ everywhere within the stellar structure. In the absence of a cosmological constant, Buchdahl \cite{Buchdahl:1959zz} proposed the upper bound for surface redshift for the isotropic general relativistic objects as $z_s~\leq~2$, whereas Boehmer and Harko \cite{Boehmer:2006ye} generalized the bound for an anisotropic spherical object in the presence of a cosmological constant as $z_s \leq 5$ which is also consistent with the maximum bound $z_s \leq 5.211$ derived by Ivanov \cite{Ivanov:2002xf}. The effective surface redshift strongly depends on the physical parameters at the center of the star.

               Finally, we can conclude that our proposed strange star model obtained by employing isotropic Durgapal's fourth solution in $f(R,\,T)$ gravity theory is singularity-free, which also satisfies all physical conditions of a stable and viable model. Therefore, this kind of solution may be suitable for modeling known strange compact objects.

\end{itemize}

\section*{Declarations}
\textbf{Funding:} The authors did not receive any funding in the form of financial aid or grant from any institution or organization for the present research work.\par
\textbf{Data Availability Statement:} The results are obtained
via purely theoretical calculations and can be verified analytically,
thus this manuscript has no associated data, or the data will not be deposited. \par
\textbf{Conflicts of Interest:} The authors have no financial interest or involvement which is relevant by any means to the content of this study.
\\ \\

\section*{Acknowledgements} P.R. and P.B. both are thankful to the Inter-University Centre for Astronomy and Astrophysics (IUCAA), Pune, Government of India, for providing Visiting Associateship.

\bibliography{ref1}

\end{document}